# A quantum encryption design featuring confusion, diffusion, and mode of operation

Zixuan Hu and Sabre Kais*

Department of Chemistry, Department of Physics, and Purdue Quantum Science and Engineering Institute, Purdue University, West Lafayette, IN 47907, United States
E-mail: kais@purdue.edu

Abstract: Quantum cryptography – the application of quantum information processing and quantum computing techniques to cryptography has been extensively investigated. Two major directions of quantum cryptography are quantum key distribution (QKD) and quantum encryption, with the former focusing on secure key distribution and the latter focusing on encryption using quantum algorithms. In contrast to the success of the QKD, the development of quantum encryption algorithms is limited to designs of mostly one-time pads (OTP) that are unsuitable for most communication needs. In this work we propose a non-OTP quantum encryption design utilizing a quantum state creation process to encrypt messages. As essentially a non-OTP quantum block cipher the method stands out against existing methods with the following features: 1. complex key-ciphertext relation (i.e. confusion) and complex plaintext-ciphertext relation (i.e. diffusion); 2. mode of operation design for practical encryption on multiple blocks. These features provide key reusability and protection against eavesdropping and standard cryptanalytic attacks.

## 1. Introduction

Cryptography – the study of secure communication in the presence of eavesdropping adversaries – is an important application of classical computing and information processing. Inspired by the rapid progress in both theory and experiment, the application of quantum computing and information processing techniques to cryptography has been extensively investigated [1-4]. A prominent example is the potential of Shor's factorization algorithm [5] to break the most widely used public-key encryption system. Facing this challenge, classical cryptography is considering post-quantum cryptographic systems [6, 7] that are secure against current and future quantum algorithms. On the other hand, the emergence of cryptographic systems based on quantum technologies has led to the burgeoning field of quantum cryptography. Currently there are two major directions of quantum cryptography: quantum key distribution (QKD) and quantum encryption algorithm. The QKD [2, 3, 8-11] focuses on secure key generation and distribution by exploiting quantum phenomena such as the probabilistic nature of quantum measurement and the non-locality of entanglement. The development of the QKD has successfully produced widely accepted key-distribution protocols such as the BB84 [3]. Note that the QKD only processes the keys while the encryption process, decryption process, and the communication process have to use established classical algorithms and channels. Quantum encryption algorithm on the other hand uses quantum computing techniques to encrypt messages (classical or quantum) into quantum states that are communicated to and decrypted by the recipient. In contrast to the well accepted success of the QKD, the development of quantum encryption algorithms is rather limited to designs [12-14] that are mostly quantum versions of the one-time pad (OTP). The OTP is an encryption scheme that ensures perfect secrecy [15] in the sense that the ciphertext (i.e. the encrypted message) provides no information at all on the plaintext (i.e. the original message) to any cryptanalytic attempt – which means the OTP is unbreakable even with infinite computational resources. However, a critical problem with using the OTP is that each original message requires a unique key of the same length as the message itself. As the key must be random and can never be re-used [15], the generation, transfer, and storage of indefinite amount of keys for an OTP are difficult in practice, making the OTP not suitable for the majority of the communication needs of the present day. Consequently most widely used encryption methods such as the symmetric encryption Advanced Encryption Standard (AES) [16] and the asymmetric encryption Rivest-Shamir-Adleman (RSA) [17] offer not perfect secrecy but practical secrecy [15] – i.e. breaking the encryption requires currently unrealistic computational resources. In this work we propose a new non-OTP quantum encryption design that utilizes a quantum state creation process to encrypt messages. Using a quantum state as the ciphertext, the quantum encryption offers an inherent level of protection against eavesdropping, because without the key any brute force measurement of the ciphertext state will collapse it into a random basis state. The non-readability of the ciphertext is a unique advantage of quantum encryption over classical methods where the ciphertext is just a bit string. Next we introduce the concepts of confusion (complex key-ciphertext relation) and diffusion (complex plaintext-ciphertext relation) from classical cryptography into quantum encryption and propose a novel encryption process that creates both confusion and diffusion. This ensures that small differences in the plaintext lead to substantial changes in the ciphertext or vice versa, such that the inability of a potential adversary to analyze the ciphertext state is amplified. Finally, we introduce the concept of mode of operation from classical cryptography into quantum encryption to enable practical encryption on arbitrary number of blocks of plaintexts. The mode of operation procedures developed for the quantum encryption design generalize the classical cipher block chaining (CBC) [18] to work with a quantum ciphertext by exploiting unique properties of quantum measurement and quantum superposition. The quantum mode of operation therefore has truly random or unreadable initialization vectors that are impossible for the classical CBC mode. The adaptation of confusion, diffusion and mode of operation from classical cryptography into quantum cryptography not only provides key reusability and stronger security against standard cryptanalytic attacks but also establishes new design principles for the systematic development of quantum encryption methods which may lead to improved quantum cryptographic systems beyond the particular design of the current study.

## 2. Theory of the quantum encryption design

### 2.1 Encrypting classical data with quantum states.

The essence of any encryption method with practical secrecy is a reversible process whose computational cost strongly depends on a secret piece of information called the key. In this work we focus on the symmetric-key scenario where decryption uses the same key as encryption. Consider an *n*-bit classical plaintext, practical secrecy is defined such that for the legitimate parties of the communication Alice and Bob knowing the key, both encryption and decryption are computationally simple in the sense that the number of computational steps required is polynomial: i.e. $O(cn^k)$ for some constant *c* and *k* such that $cn^k$ is overwhelmingly smaller than $2^n$. In the meanwhile, for the adversary Eve not knowing the key, both encryption and decryption are computationally hard in the sense that the number of computational steps required is exponential: i.e. much greater than $O(2^n)$. To achieve this with quantum encryption Alice starts with an *n*-qubit quantum state in the initial state $|0\rangle^{\otimes n}$. The first step Alice applies at most *n* Pauli-X gates to encode an *n*-bit classical plaintext into a quantum state plaintext: e.g. 00101 is coded into $|00101\rangle$. The second step she applies a polynomial sequence of 1-qubit and 2-qubit elementary gates to transform the quantum plaintext into a quantum state that serves as the quantum ciphertext, and then sends it to Bob. The account of the polynomial sequence of elementary gates used by Alice is the key pre-shared with Bob such that upon receiving the quantum ciphertext Bob can apply the inverse operations to recover the quantum plaintext. The classical plaintext can then be revealed by projection measurement on the quantum plaintext in the computational basis. So far without going into any detail of the encryption procedure, the just described process is not so different from a generalization of existing studies of quantum encryption [12-14, 19], and we will later in Sections 2.2 and 2.3 present the new quantum encryption design with confusion, diffusion, and mode of operation that provide key reusability and stronger security. However, here we first discuss certain security already provided by just considering the quantum nature of the ciphertext.

Firstly, note the fact that a quantum state ciphertext naturally contains more uncertainty than a classical ciphertext. For example a classical bit 0 (1) can be mapped to a qubit state $|0\rangle$ ($|1\rangle$), which after a unitary operation becomes $a_1|0\rangle + a_2|1\rangle$ ($a_2^*|0\rangle - a_1^*|1\rangle$), where $|a_1|^2 + |a_2|^2 = 1$. For encryption purpose a ciphertext in the form of $a_1|0\rangle + a_2|1\rangle$ presents more difficulty to the eavesdropper Eve, because even if she has successfully intercepted the state $a_1|0\rangle + a_2|1\rangle$, without the key (i.e. the value of $a_1$) she cannot reliably read the content of the ciphertext. In practice if we assume $a_1$ can take $N$ discrete values between 0 and 1, the uncertainty associated with it is typically far greater than 1 bit as $N \gg 2$. This difficulty for Eve is much more significant for a multi-qubit ciphertext state in which qubits are entangled with each other. This is because a brute-force measurement on the ciphertext state destroys the intricate dependencies among qubits and collapses the ciphertext into a simple state with all qubits in either $|0\rangle$ or $|1\rangle$: such a state has little resemblance to either the ciphertext state or the plaintext. Consequently quantum encryption exploits the quantum phenomena of superposition and entanglement to produce a ciphertext that cannot even be read without the key. In comparison, a classical ciphertext is typically a bit-string with the same length as the plaintext, and it can be read and analyzed by Eve to gain information on the key and the plaintext.

Secondly, even if Eve is able to read the ciphertext – assuming the rare and can-be-avoided scenario that Alice sends the same ciphertext state many times and Eve is able to gain statistical knowledge of it – it is still highly difficult for her to deduce the key or the plaintext from the ciphertext. The detail of this reasoning is presented in the Supplementary Information where the quantum state complexity theory in our previous study has been used [20]. Furthermore, this compromising scenario of Alice sending the same copy of the ciphertext many times can be totally avoided by the confusion, diffusion, and mode of operation to be introduced in the following sections.

**2.2 The quantum encryption with confusion and diffusion.**

So far we have seen two security features by using a quantum state as the ciphertext: the difficulty in reading the quantum ciphertext and the impossibility to deduce the key even if the quantum ciphertext is somehow known. These features however are not sufficient for a good encryption method: to provide reusability of keys and protection against standard cryptanalytic attacks we need to design an encryption with good confusion and diffusion [15]. Confusion means complex relation between the ciphertext and the key such that it is difficult to deduce key properties by analyzing the patterns in ciphertexts. Classically if one bit in the ciphertext depends on multiple parts of the key, confusion is provided. For our quantum encryption design, as the ciphertext cannot be measured deterministically, confusion can be accordingly defined that the statistics of measuring one qubit in the ciphertext state depends on multiple parts of the key. Diffusion means complex relation between the plaintext and the ciphertext such that it is difficult to deduce plaintext properties by analyzing the patterns in ciphertexts or vice versa. Classically if changing one bit in the plaintext (ciphertext) changes more than half of the bits in the ciphertext (plaintext), diffusion is provided. Again since in our quantum encryption the ciphertext cannot be measured deterministically, diffusion can be defined that changing the value of one qubit in the plaintext leads to changes of statistics of measuring more than half of the qubits in the ciphertext. Note the vice versa ciphertext-to-plaintext relation is not defined for the quantum case because it is impossible to create a proper ciphertext without knowing the plaintext and the key first.

We start with a basic encryption design where one unitary $U_i$ with real parameters (for simplicity we assume all parameters in the following discussions are real, however the method can be generalized to have complex parameters) is applied to each qubit $q_i$ of the plaintext, and no CNOT is applied. The key is then the collection $\{U_i\}$ where the order of $U_i$'s is unimportant. Clearly this encryption does not provide either confusion or diffusion because the statistical pattern of measuring each qubit $q_i$ of the ciphertext depends on only one part of the key $U_i$ and only one qubit (the same $q_i$) of the plaintext. For example suppose after this step in the ciphertext $q_1 = a_1|0\rangle_1 + a_2|1\rangle_1$ and $q_2 = b_1|0\rangle_2 + b_2|1\rangle_2$, then the probability of measuring $|0\rangle$ for $q_1$ is $p(|0\rangle_1) = a_1^2$ and the probability of measuring $|0\rangle$ for $q_2$ is $p(|0\rangle_2) = b_1^2$. If this key is reused many times, Eve would be able to deduce $U_1$ and $U_2$ by measuring the probability of outcomes for $q_1$ and $q_2$ of the ciphertext (the same for all other qubits). Now after this step if we apply $\text{CNOT}_{1\to 2}$ (where $1 \to 2$ means $q_1$ is the control and $q_2$ is the target), the 2-qubit state is:

$$\phi^{(2)} = a_1|0\rangle_1 \left(b_1|0\rangle_2 + b_2|1\rangle_2\right) + a_2|1\rangle_1 \left(b_1|1\rangle_2 + b_2|0\rangle_2\right) \tag{1}$$

then by simple calculation $p(|0\rangle_1) = a_1^2$ still but $p(|0\rangle_2) = a_1^2 b_1^2 + a_2^2 b_2^2$ – we see that $q_2$ gains a dependence on $U_1$ in the sense that the probabilities of outcomes when measuring $q_2$ depend on $U_1$ after $\text{CNOT}_{1\to 2}$ is applied. If we further apply $\text{CNOT}_{2\to 3}$ to $q_3 = c_1|0\rangle_3 + c_2|1\rangle_3$, the 3-qubit state is:

$$\phi^{(3)} = \left(a_1 b_1|0\rangle_1 + a_2 b_2|1\rangle_1\right)|0\rangle_2 \left(c_1|0\rangle_3 + c_2|1\rangle_3\right) + \left(a_1 b_2|0\rangle_1 + a_2 b_1|1\rangle_1\right)|1\rangle_2 \left(c_1|1\rangle_3 + c_2|0\rangle_3\right) \tag{2}$$

then $p(|0\rangle_1) = a_1^2$, $p(|0\rangle_2) = a_1^2 b_1^2 + a_2^2 b_2^2$, $p(|0\rangle_3) = \left(a_1^2 b_1^2 + a_2^2 b_2^2\right)c_1^2 + \left(a_1^2 b_2^2 + a_2^2 b_1^2\right)c_2^2$ – i.e. $q_3$ gains dependences on both $U_1$ and $U_2$. The results in Eqs. (1) and (2) reveal the effects of 1-qubit unitaries and CNOT's from a cryptographic perspective:

> **Theorem 1.** If the probabilities of outcomes when measuring a qubit depend on some 1-qubit unitaries applied to this or any other qubit, we say this qubit has dependences on these 1-qubit unitaries. Then a 1-qubit unitary creates dependences on its target qubit and a CNOT causes the target qubit to gain all the dependences from the control qubit, while the control qubit retaining all its dependences.

**Proof of Theorem 1.** Suppose $q_1$ is one qubit in a general $n$-qubit state $\phi^{(n)}$, the Schmidt decomposition of $\phi^{(n)}$ with respect to $q_1$ is:

$$\phi^{(n)} = C_1 \phi_1^{(n-1)} \left( a_1 |0\rangle_1 + a_2 |1\rangle_1 \right) + C_2 \phi_2^{(n-1)} \left( a_2 |0\rangle_1 - a_1 |1\rangle_1 \right) \tag{3}$$

where $\phi_1^{(n-1)}$ and $\phi_2^{(n-1)}$ are orthogonal, and therefore $p(|0\rangle_1) = C_1^2 a_1^2 + C_2^2 a_2^2$: this means $q_1$ depends on the pairs $(C_1, C_2)$ and $(a_1, a_2)$ that are created by previous quantum operations used to generate $\phi^{(n)}$. Now applying another unitary gate $U = \begin{pmatrix} u_1 & u_2 \\ u_2 & -u_1 \end{pmatrix}$ to $q_1$ we get:

$$\begin{aligned} U\phi^{(n)} &= C_1 \phi_1^{(n-1)} \left[ (a_1 u_1 + a_2 u_2) |0\rangle_1 + (a_1 u_2 - a_2 u_1) |1\rangle_1 \right] \\ &+ C_2 \phi_2^{(n-1)} \left[ (a_2 u_1 - a_1 u_2) |0\rangle_1 + (a_2 u_2 + a_1 u_1) |1\rangle_1 \right] \end{aligned} \tag{4}$$

where $p(|0\rangle_1) = C_1^2 (a_1 u_1 + a_2 u_2)^2 + C_2^2 (a_2 u_1 - a_1 u_2)^2$, so indeed $q_1$ has gained dependence on $U$. Note that for any $U$, $(a_1 u_1 + a_2 u_2)|0\rangle_1 + (a_1 u_2 - a_2 u_1)|1\rangle_1$ is always orthogonal to $(a_2 u_1 - a_1 u_2)|0\rangle_1 + (a_2 u_2 + a_1 u_1)|1\rangle_1$, and thus the probabilities of no qubit other than $q_1$ are affected by $U$. Now suppose we further Schmidt-decompose $\phi_1^{(n-1)}$ and $\phi_2^{(n-1)}$ in Equation (3) with respect to another qubit $q_2$:

$$\begin{aligned} \phi^{(n)} &= C_1 \left[ D_{11} \phi_{11}^{(n-2)} \left( b_{11} |0\rangle_2 + b_{12} |1\rangle_2 \right) + D_{12} \phi_{12}^{(n-2)} \left( b_{12} |0\rangle_2 - b_{11} |1\rangle_2 \right) \right] \left( a_1 |0\rangle_1 + a_2 |1\rangle_1 \right) \\ &+ C_2 \left[ D_{21} \phi_{21}^{(n-2)} \left( b_{21} |0\rangle_2 + b_{22} |1\rangle_2 \right) + D_{22} \phi_{22}^{(n-2)} \left( b_{22} |0\rangle_2 - b_{21} |1\rangle_2 \right) \right] \left( a_2 |0\rangle_1 - a_1 |1\rangle_1 \right) \end{aligned} \tag{5}$$

where $\langle \phi_{11}^{(n-2)} | \phi_{12}^{(n-2)} \rangle = \langle \phi_{21}^{(n-2)} | \phi_{22}^{(n-2)} \rangle = 0$, and then we can calculate the probability: $p(|0\rangle_2) = C_1^2 (D_{11}^2 b_{11}^2 + D_{12}^2 b_{12}^2) + C_2^2 (D_{21}^2 b_{21}^2 + D_{22}^2 b_{22}^2)$. We see that $q_1$ and $q_2$ share a dependence on the pair $(C_1, C_2)$ but the dependence on $(a_1, a_2)$ is unique to $q_1$. Now apply $\text{CNOT}_{1 \to 2}$ to $\phi^{(n)}$:

$$\begin{aligned} \text{CNOT}_{1 \to 2} \phi^{(n)} &= \begin{pmatrix} a_1 C_1 \left[ D_{11} \phi_{11}^{(n-2)} \left( b_{11} |0\rangle_2 + b_{12} |1\rangle_2 \right) + D_{12} \phi_{12}^{(n-2)} \left( b_{12} |0\rangle_2 - b_{11} |1\rangle_2 \right) \right] \\ + a_2 C_2 \left[ D_{21} \phi_{21}^{(n-2)} \left( b_{21} |0\rangle_2 + b_{22} |1\rangle_2 \right) + D_{22} \phi_{22}^{(n-2)} \left( b_{22} |0\rangle_2 - b_{21} |1\rangle_2 \right) \right] \end{pmatrix} |0\rangle_1 \\ &+ \begin{pmatrix} a_2 C_1 \left[ D_{11} \phi_{11}^{(n-2)} \left( b_{11} |1\rangle_2 + b_{12} |0\rangle_2 \right) + D_{12} \phi_{12}^{(n-2)} \left( b_{12} |1\rangle_2 - b_{11} |0\rangle_2 \right) \right] \\ - a_1 C_2 \left[ D_{21} \phi_{21}^{(n-2)} \left( b_{21} |1\rangle_2 + b_{22} |0\rangle_2 \right) + D_{22} \phi_{22}^{(n-2)} \left( b_{22} |1\rangle_2 - b_{21} |0\rangle_2 \right) \right] \end{pmatrix} |1\rangle_1 \end{aligned} \tag{6}$$

After some algebra we obtain:

$$p(|0\rangle_2) = a_1^2 \left[ C_1^2 \left( D_{11}^2 b_{11}^2 + D_{12}^2 b_{12}^2 \right) + C_2^2 \left( D_{21}^2 b_{22}^2 + D_{22}^2 b_{21}^2 \right) \right] + a_2^2 \left[ C_2^2 \left( D_{21}^2 b_{21}^2 + D_{22}^2 b_{22}^2 \right) + C_1^2 \left( D_{11}^2 b_{12}^2 + D_{12}^2 b_{11}^2 \right) \right]$$

$$+ 2a_1 a_2 C_1 C_2 \begin{bmatrix} D_{11} D_{21} \langle \phi_{11}^{(n-2)} | \phi_{21}^{(n-2)} \rangle (b_{11} b_{21} - b_{12} b_{22}) + D_{11} D_{22} \langle \phi_{11}^{(n-2)} | \phi_{22}^{(n-2)} \rangle (b_{11} b_{22} + b_{12} b_{21}) \\ + D_{12} D_{21} \langle \phi_{12}^{(n-2)} | \phi_{21}^{(n-2)} \rangle (b_{12} b_{21} + b_{11} b_{22}) + D_{12} D_{22} \langle \phi_{12}^{(n-2)} | \phi_{22}^{(n-2)} \rangle (b_{12} b_{22} - b_{11} b_{21}) \end{bmatrix}$$

(7)

where we see that $q_2$ has gained dependence on the pair $(a_1, a_2)$, which was originally unique to $q_1$. Because the form of $\phi^{(n)}$ in Equation (3) is entirely general, $q_1$'s dependence on $(a_1, a_2)$ can be understood as a package including all its dependences gained in the process of creating $\phi^{(n)}$ – through either 1-qubit unitaries applied to $q_1$ or CNOT's applied to $q_1$ as the target. Equation (7) shows that by a single $\text{CNOT}_{1\to 2}$ all $q_1$'s dependences packaged in $(a_1, a_2)$ are created on $q_2$. It is trivial to see that $q_1$ still retains its dependences. This concludes the proof for Theorem 1. Note that the dependences created on $q_2$ are not the same as those on $q_1$ – the probabilities indeed depend on the same unitaries, but the exact forms are different. Theorem 1 is significant that it allows us to create new probability dependences with 1-qubit unitaries on selective qubits and then efficiently pass them onto other qubits by CNOT gates. In the following we show how to use this result to design an encrypting process with good confusion and diffusion properties.

**The encrypting process with good confusion and diffusion:**

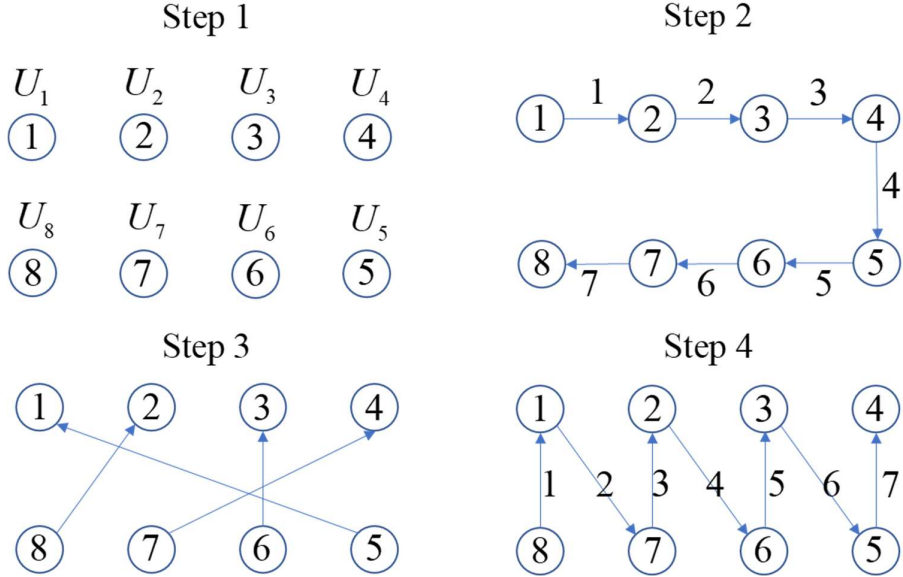

Figure 1. Graphical illustration of the encrypting process with an 8-qubit example. The circles with numbers inside represent the qubits. The arrows represent CNOT gates for which each arrow begins at the control qubit and points to the target qubit. The numbers on the arrows indicate the order in which the CNOT gates are applied within the current step. Step 1: apply a 1-qubit $U_i$ to each qubit $q_i$. Step 2: apply $\text{CNOT}_{i\to i+1}$ sequentially for $i = 1$ to $n-1$, this step causes the downstream qubits 5 to 8 to gain dependences on more than half of the $U_i$'s. Step 3: use the downstream qubits 5 to 8 as controls and the upstream qubits 1 to 4 as targets to apply

CNOT gates. Showing one example out of the $\left(\frac{n}{2}\right)!$ possible ways the qubits are paired. The CNOT gates in this step all commute so the order is unimportant. After this step confusion is achieved. Step 4: with the general goal of achieving diffusion, this step has great freedom. In the particular example shown here, a series of CNOT gates run alternately between the downstream and upstream qubits. After this step diffusion is achieved.

Start with an *n*-qubit plaintext where each qubit $q_i$ is either $|0\rangle$ or $|1\rangle$.

**Step 1**: Apply a 1-qubit unitary $U_i$ to each qubit $q_i$ and create the initial dependence of each $q_i$ to its corresponding $U_i$. This is the basic key design mentioned earlier. If each $U_i$ is defined by a real parameter that can take *N* discrete values, there are totally $N^n$ possibilities that contribute to key size. This step costs $n$ $U_i$ gates.

**Step 2**: Apply $CNOT_{i \to i+1}$ sequentially for $i = 1$ to $n-1$: i.e. $CNOT_{1 \to 2}$ first, then $CNOT_{2 \to 3}$, then $CNOT_{3 \to 4}$, ... , finally $CNOT_{n-1 \to n}$. By Theorem 1, the $CNOT_{1 \to 2}$ causes $q_2$ to gain the dependence on $U_1$ from $q_1$, and then $CNOT_{2 \to 3}$ causes $q_3$ to gain all the dependences from $q_2$ that include both $U_2$ from $q_2$ itself and $U_1$ that $q_2$ has just gained from $q_1$. In such a snowball process, each further $CNOT_{k \to k+1}$ causes $q_{k+1}$ to gain dependences on all the $U_i$'s for $i \leq k$. After this step each $q_i$ with $i > \frac{n}{2}$ has gained dependences on more than half of the $U_i$'s. We remark that the order of the application of the $CNOT_{i \to i+1}$ gates is important: if we apply $CNOT_{2 \to 3}$ before $CNOT_{1 \to 2}$, $q_2$ has not gained the dependence on $U_1$ from $q_1$ yet and thus $q_3$ will not gain that dependence either. Applying $CNOT_{2 \to 3}$ before $CNOT_{1 \to 2}$ is therefore less efficient than applying $CNOT_{2 \to 3}$ after $CNOT_{1 \to 2}$ as the latter can pass more dependences from $q_2$ to $q_3$. This step costs $n-1$ CNOT gates.

**Step 3**: For each $q_i$ with $i > \frac{n}{2}$ (the downstream qubits), randomly assign a different $q_k$ with $k \leq \frac{n}{2}$ (the upstream qubits), such that all the qubits are paired (except one qubit if *n* is odd). Apply $CNOT_{i \to k}$ for each pair such that the upstream $q_k$ gains all the dependences from the downstream $q_i$. After Step 2 each downstream $q_i$ with $i > \frac{n}{2}$ depends on more than half of the $U_i$'s, and here by the $CNOT_{i \to k}$ gates these downstream qubits pass all their dependences to the corresponding upstream qubits. Consequently now each one of the *n* qubits will have gained dependences on more than half of the $U_i$', and this complex relation between the ciphertext and the key provides confusion as defined earlier. The process that gets all qubits into pairs has $\left(\frac{n}{2}\right)!$

possibilities that contribute to key size (minor changes if *n* is odd). This step costs $\frac{n}{2}$ CNOT gates.

**Step 4**: Now to achieve diffusion defined earlier we want the property that changing the value of one qubit in the plaintext changes the statistics of measuring more than half of the qubits in the ciphertext. Suppose a qubit $q_j$ is $|0\rangle$ in the plaintext, after $U_j$ in Step 1 it becomes $a_1|0\rangle_j + a_2|1\rangle_j$ and $p(|0\rangle_1) = a_1^2$. If the plaintext $q_j$ is changed to $|1\rangle$ then after $U_j$ it becomes $a_2|0\rangle_j - a_1|1\rangle_j$ and $p(|0\rangle_1) = a_2^2$, so the dependence of $q_j$ on $U_j$ has changed. In addition, although the minus sign in $a_2|0\rangle_j - a_1|1\rangle_j$ does not immediately have an effect on probabilities, it can change how the subsequent qubits depend on $U_j$ after Steps 2 and 3. Hence we see that a value change in one qubit $q_j$ in the plaintext will affect all the ciphertext qubits that have gained dependences from $q_j$.

This means that any upstream qubit $q_k$ with $k \leq \frac{n}{2}$ already has diffusion after Steps 2, because all the downstream qubits in the ciphertext (more than half of all qubits) have gained dependences from $q_k$. Now to create diffusion in the downstream qubits, we just need to use these qubits as control and apply CNOT gates to random qubits as targets (can be either upstream or downstream) until on average more than half of all qubits have gained dependences from any qubit. For example, two qubits have gained dependences from the last qubit $q_n$ after Step 3: $q_n$ itself and the qubit assigned to pair with $q_n$, thus we need to apply at most $\frac{n}{2} - 2$ CNOT gates using $q_n$ as the control to pass $q_n$'s dependences to half of all qubits. The actual CNOT gates required may be fewer than $\frac{n}{2} - 2$ because we can first pass $q_n$'s dependences to another downstream qubit such as $q_{n-2}$, and then any CNOT gate using $q_{n-2}$ as the control will also pass $q_n$'s dependences to the target. In fact, an example of a very efficient implementation is to apply a series of CNOT gates running alternately through the downstream and upstream qubits, where the target qubit of the previous CNOT serves as the control qubit of the next CNOT: e.g. $\text{CNOT}_{n \to 1}$ first, then $\text{CNOT}_{1 \to n-1}$, then $\text{CNOT}_{n-1 \to 2}$, then $\text{CNOT}_{2 \to n-2}$, ... , finally $\text{CNOT}_{n/2 \to n/2+1}$. By Theorem 1 it is easy to verify that this implementation guarantees more than half of all qubits have gained dependences from any downstream qubit. Unlike the previous steps, Step 4 allows greater freedom in the key design and the exact evaluation of the contribution to key size and gate cost is impossible. However, for the particular implementation just described, the order of the upstream qubits can be any permutation and thus there are $\left(\frac{n}{2}\right)!$ possibilities that contribute to key size. This implementation costs *n* CNOT gates.

Step 4 concludes the ciphertext creation process. A graphical illustration of the four steps of encryption is drawn in Figure 1. The account of all the unitaries and CNOT gates used is the key shared with the recipient, who can then recover the plaintext by reversing all the gates.

Through the description and analysis of the encrypting process, we can see that our quantum encryption design supports efficient implementation with $O(n)$ gates and large key size with at least $O\left(N^n\left(\frac{n}{2}\right)!\right)$ possible variations. More importantly the design has provable confusion and diffusion that makes the key reusable while protecting against common cryptanalytic attacks.

**2.3 Mode of operation.**

The quantum encryption described so far is a block cipher where each block of message containing $n$ bits of classical information is encrypted into a quantum state of $n$ qubits. Similar to the classical counterpart, the quantum block cipher also requires a mode of operation to ensure that different ciphertexts (blocks) are generated even with the same plaintext and key used. This feature together with diffusion allows the key to be reused many times to securely transmit large amount of information. Our mode of operation is inspired by the classical cipher block chaining (CBC) [18]. In the CBC mode a randomly chosen $n$-bit initialization vector (IV) is XORed ($\oplus$) with the plaintext $P_1$ of the first block, the encrypting algorithm then works on $\text{IV} \oplus P_1$ to produce the first ciphertext $C_1$. Next $C_1$ is XORed with the plaintext $P_2$ of the second block before it is encrypted into $C_2$. Repeat this process many times where each time the plaintext $P_i$ of the current block is XORed with the ciphertext $C_{i-1}$ of the previous block before getting encrypted into the ciphertext $C_i$ of the current block:

$$C_i = E(K)(P_i \oplus C_{i-1}), \quad C_0 = \text{IV} \tag{8}$$

where $E(K)$ is the encrypting function with the key $K$. To generalize the CBC to our quantum encryption, the ciphertext here is a quantum state that cannot be directly XORed with the plaintext of the following block, and in the following we propose two different modes to solve this problem.

In the first mode shown in Figure 2, after the first ciphertext state $|C_1\rangle$ has been created $|C_1\rangle = E(K)(P_1 \oplus \text{IV})$, we create an additional copy of $|C_1\rangle$ and measure it in the computational basis $\{|0\rangle, |1\rangle\}$. Because the measurement result has every qubit in $|0\rangle$ or $|1\rangle$, it can be used as the new classical IV to be XORed with the plaintext of the following block. Repeat this process iteratively:

$$|C_i\rangle = E(K)(P_i \oplus M(|C_{i-1}\rangle)), \quad M(|C_0\rangle) = \text{IV} \tag{9}$$

where $M(|C_{i-1}\rangle)$ is the measurement result on the copy of $|C_{i-1}\rangle$.

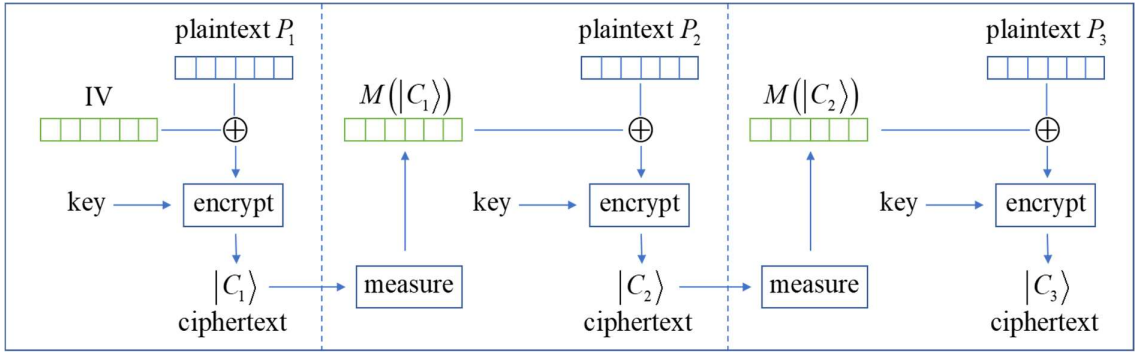

Figure 2. The first mode of operation mechanism shown with a 3-block example. In each iteration after the first one, the ciphertext state $|C_{i-1}\rangle$ is measured into a classical IV that is then XORed with the plaintext $P_i$.

In the second mode shown in Figure 3, after the first ciphertext state has been created by $C_1 = E(K)(P_1 \oplus \text{IV})$, we use the qubits of $C_1$ as controls to apply CNOT gates to the qubits of the following plaintext. Each qubit on $C_1$ as the control is paired with a different qubit on the following plaintext as the target. For simplicity, the same pairing plan that specifies which qubit of the current ciphertext state controls which target qubit of the next plaintext can be used for each iteration. Repeat this process iteratively:

$$|C_1\rangle = E(K)(P_1 \oplus \text{IV}),$$
$$|C_i\rangle = E(K)\left(\text{CNOT}(|C_{i-1}\rangle \to |P\rangle_i)\right), \quad i > 1 \tag{10}$$

where in the process $\text{CNOT}(|C_{i-1}\rangle \to |P\rangle_i)$ each qubit on the ciphertext state $|C_{i-1}\rangle$ as the control applies a CNOT to a different qubit on the plaintext state $|P\rangle_i$ as the target.

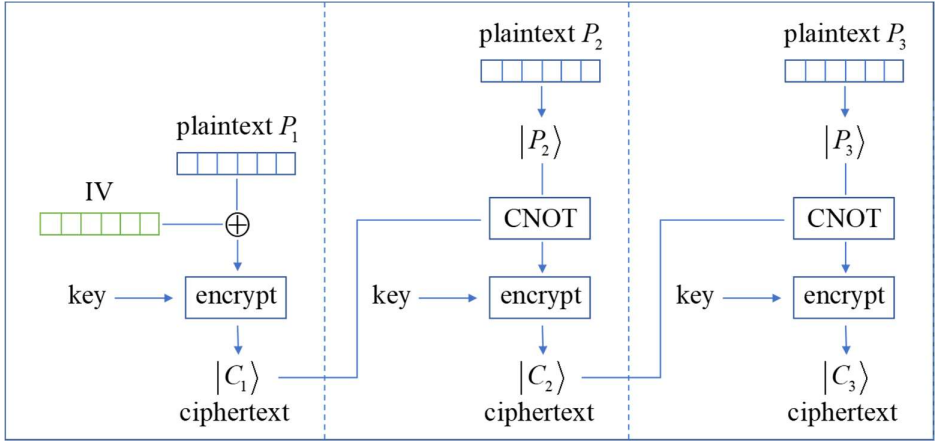

Figure 3. The second mode of operation mechanism shown with a 3-block example. In each iteration after the first one, each qubit on the ciphertext state $|C_{i-1}\rangle$ as the control applies a CNOT to a different qubit on the plaintext state $|P\rangle_i$ as the target.

Compared to the classical CBC, both quantum modes of operation have additional security because the IV for each iteration is not simply the ciphertext of the last block that is revealed to Eve. For the first mode, the IV for each iteration is generated with the truely random process of quantum measurement on the previous ciphertext state. For the second mode, all IV's after the first one are quantum states that cannot be reliably read. Furthermore, in the second mode the pairing plan of which qubit on the IV controls which qubit on the next plaintext can be pre-shared as additional parts of the key – which has $n!$ complexity. Both quantum modes of operation ensure different ciphertexts are generated even with the same plaintext and key used.

### 2.4 Discussions.

The mode of operation together with the encryption process completes our description of the new quantum encryption design. In actual application, Alice will first encode the classical bit string into a quantum basis state (e.g. 00101 is coded into $|00101\rangle$), and then apply a sequence of quantum gates following the procedure in Section 2.2 to create a quantum ciphertext. Note that the procedure in Section 2.2 is only a guideline to ensure confusion and diffusion by the result of Theorem 1. In this sense Theorem 1 can be considered as a foundational result that may inspire many other encryption procedures in addition to the particular one described in this work. Nonetheless the procedure in Section 2.2 already provides great freedom with at least $O\left(N^n \left(\frac{n}{2}\right)!\right)$ variations contributing to the key size if a brute force attack is attempted. On the other hand the implementation cost of the procedure is only $O(n)$ gates, which is very efficient.

The ciphertext state can then be sent to Bob through an unsecure channel with possible eavesdropping by Eve. An account of the exact sequence of quantum gates applied by Alice is the key shared with Bob through a secure channel – note this can be done long before the actual communication happens thus it is harder to expect and attack by Eve. Upon receipt of the ciphertext state, Bob can apply the inverse quantum operations to recover the plaintext. After the first block of plaintext, additional blocks of plaintexts can be encrypted with additional mode of operation procedures as described in Section 2.3 such that the statistics of the ciphertext state is further disguised.

The security of the quantum encryption design is provided by multiple mechanisms. Firstly the use of a quantum state as the ciphertext makes it impossible for Eve to reliably read and analyze the ciphertext. This is a unique quantum advantage over classical methods for which the ciphertext is just a bit string. In principle Eve could gain statistical knowledge of the ciphertext if the same one is sent many times, but this possibility is prevented by implementing one of the two quantum modes of operation. The two quantum modes of operation provide truly random or unreadable initialization vectors depending on the mode of choice, and these are impossible for classical modes of operation. Having provable confusion and diffusion provides our method an additional layer of protection against potential cryptanalysis, because small changes in the plaintext lead to substantial changes in the ciphertext or vice versa. On the contrary, knowing the key, the legitimate recipient Bob can easily reverse the encrypting process to generate the plaintext deterministically from the ciphertext, without the need to actually read the ciphertext. The unique situation that the ciphertext can lead to the plaintext deterministically while not readable itself, together with features like confusion, diffusion, and mode of operation, make our quantum encryption strongly resistant to cryptanalytic attacks. For example, the chosen-plaintext attack (CPA) and the chosen-ciphertext attacks (CCA1 and CCA2) require Eve to analyze a few plaintext-ciphertext pairs to gain knowledge of the key. Now that the ciphertext being unreadable, and the statistics being obscured by confusion, diffusion, and mode of operation, it is very difficult for Eve to extract information from a few plaintext-ciphertext pairs. In addition, eavesdropping by Eve on the ciphertext inevitably disturbs the quantum state such that the recipient Bob can detect such interception. For Bob to determine if his measurement result is the correct message, the message disturbed by Eve, or the message corrupted by inherent system uncertainties (gate error, channel noise, etc.), multiple blocks of the same plaintext should be sent thus to establish a protocol analogous to the repetition code for error correcting purposes. As an interesting idea for future studies, the exact number of repetitions required for reliable communication should depend on the gate quality, channel quality, and key design.

## 3. Conclusion

In this work we have developed a quantum encryption design that utilizes a quantum state creation process to encrypt messages. By using a quantum state as the ciphertext and the creation procedure as the key, an inherent level of security is guaranteed by the statistical nature of quantum measurements as well as the complexity of the state creation process. We then introduce the concepts of confusion and diffusion from classical cryptography into quantum encryption and provide both features with a novel quantum encryption process. Finally we introduce the concept of mode of operation from classical cryptography into quantum encryption by proposing two modes of operation inspired by the classical CBC mode. The adaptation of confusion, diffusion and mode of operation from classical cryptography into quantum cryptography not only provides key reusability and stronger security against standard cryptanalytic attacks but also establishes new design principles for the systematic development of quantum encryption methods which may lead to improved quantum cryptographic systems beyond the particular design of the current study.

**Acknowledgement**
The authors would like to acknowledge funding by the U.S. Department of Energy (Office of Basic Energy Sciences) under Award No. DE-SC0019215.

# Supplementary information: A quantum encryption design featuring confusion, diffusion, and mode of operation

Zixuan Hu and Sabre Kais*

Department of Chemistry, Department of Physics, and Purdue Quantum Science and Engineering Institute, Purdue University, West Lafayette, IN 47907, United States
E-mail: kais@purdue.eduIn the main text we stated that even if the rare and can-be-avoided scenario of Alice sending the same copy of the ciphertext many times does happen, and the adversary Eve can gain statistical knowledge of the ciphertext state, it will still be highly difficult for her to deduce the key and the plaintext. In the following we formalize this statement and present the details of the reasoning.

> **Statement 1:** For any $n$-qubit ciphertext created by applying a polynomially long sequence $K$ of 1-qubit and 2-qubit elementary gates on some $n$-qubit plaintext, suppose an adversary Eve can retrieve the coefficient associated with any basis state (e.g. she calls the basis state $|01010\rangle$ for a 5-qubit ciphertext, and get the coefficient $C_{01010}$), then she cannot deduce $K$ within polynomial number of steps.

To understand Statement 1 we first cite the result from our previous study on quantum state complexity [20] that any sequence of 1-qubit and 2-qubit elementary gates is equivalent to a sequence of 2-qubit controlled-unitary gates or $C(U)$'s. All quantum states that can be created by polynomially long sequences of 1-qubit and 2-qubit elementary gates (or concisely all polynomial states) thus correspond to all the sequences of $C(U_i)$'s with the lengths smaller than $cn^k$ for some constant $c$ and $k$ such that $cn^k$ is overwhelmingly smaller than $2^n$. The complexity of a sequence of $C(U_i)$'s comes from the parameters used to define the $U_i$'s and the *configuration* of the sequence: each $C(U_i)$ has a control qubit and a target qubit that are selected from the $n$ qubits, and the configuration of a sequence of length $cn^k$ is specified by the $cn^k$ number of control-target qubit pairs. Now suppose we are given a quantum state known to be created by a sequence of the length equal to $cn^k$, to determine the exact sequence used we need to first determine its configuration, and then the parameters of each $U_i$ can be determined by a system of equations defined by the configuration. The total possible number of configurations is $P(n,2)^{cn^k} = (n^2 - n)^{cn^k}$ (where $P(n,2)$ is the permutation of choosing one control and one target out of $n$ qubits), which is an extremely large number $\gg 2^n$. If we consider the fact that the $C(U_i)$'s may commute, then the number of unique configurations will be smaller. It is hard to evaluate the effect of commutation without some knowledge of the $C(U_i)$'s in the sequence. However, in a particular example, we can design the sequence in a way that each $C(U_i)$ is non-commutative to the previous $C(U_{i-1})$ and thus the number of unique configurations can be easily obtained. We assume the starting quantum state is created from $|0\rangle^{\otimes n}$ by applying a $U_i$ to each $q_i$ such that it is in a superposition between $|0\rangle$ and $|1\rangle$.

We apply the sequence $\left(C(U_1)_{j\to k}, C(U_2)_{k\to l}, C(U_3)_{l\to h}, C(U_4)_{h\to g}, ...\right)$ to the starting state, where the subscript $j\to k$ means $q_j$ is the control and $q_k$ is the target. In this sequence each $C(U_i)$ uses the target of the previous $C(U_{i-1})$ as the control, which ensures that $C(U_i)$ is non-commutative to $C(U_{i-1})$. As we can choose the initial $q_j$ from $n$ qubits, the initial target $q_k$ from $n-1$ qubits (excluding $q_j$), and subsequent targets from $n-1$ qubits (excluding its own control), the total number of unique configurations is $n(n-1)^{cn^k}$. This example gives a lower bound on the number of unique configurations by restricting the control of each $C(U_i)$ to be the target of the previous $C(U_{i-1})$. If we remove this restriction and consider commutation, then the actual possible number $M$ of unique configurations falls in the range $n(n-1)^{cn^k} < M < (n^2-n)^{cn^k}$: clearly $M$ is an extremely large number $\gg 2^n$. Consequently it is impossible to determine the configuration within polynomial number of steps unless there is a very efficient way to sort through the extremely large number of possible configurations. Currently there is no efficient way to relate the coefficients of a quantum state to the configuration. Although we cannot decisively prove that there will never be a way to do so – indeed it is perhaps impossible to prove that something unspecified can never happen in the future – it is highly unlikely for the following reason. Firstly as the configuration contains the information on the control and target qubits of each $C(U_i)$, it also tells us how many $C(U_i)$ gates are used to create the state. Suppose such a method is developed such that Eve could determine the configuration of any polynomial quantum state within polynomial steps, then there must exist a collection of polynomial number of procedures (each of the procedures takes polynomial steps to perform) that she can perform on an arbitrary polynomial state and discover the configuration before the procedures are exhausted. This means that given a general $n$-qubit state, she can just perform these procedures on the state assuming as if it is polynomial, and if it gives a configuration before the procedures are exhausted, then we know how many steps are required to create the state; otherwise if it does not give a configuration before the procedures are exhausted, then it is not a polynomial state. Consequently the ability of Eve to efficiently determine the configuration of a polynomial state leads to her ability to tell if a general quantum state is polynomial or not. This is a contradiction to the result proved in our previous work [20] that it is exponentially hard to determine if a general quantum state can be created within polynomial number of gates. We therefore conclude that the key of our quantum encryption design is secure even if Eve has gained significant information on the ciphertext state. Furthermore, this compromising scenario of Alice sending the same copy of the ciphertext many times can be totally avoided by the confusion, diffusion, and mode of operation introduced in the main text.